\documentclass[aps,prl,twocolumn,groupedaddress,floats,showpacs]{revtex4}
\usepackage{latexsym}
\usepackage{dcolumn}
\usepackage[dvips]{graphicx}
\usepackage{amssymb}
\usepackage{graphics}
\usepackage{amsmath}
\usepackage{epsf}

\newcommand{\vk}{{\bf k}}

\newcommand{\beq}    {\begin{equation}}
\newcommand{\enq}    {\end{equation}}

\newcommand{\rr}     {{\bf r}}

\begin{document}
\title{Theory of carrier transport in
  bilayer graphene}
\author{S. Das Sarma, E. H. Hwang, and E. Rossi} 
\affiliation{Condensed Matter Theory Center, Department of 
        Physics, University of Maryland, College Park, MD 20742-4111}

\begin{abstract}
We develop a theory for density, disorder, and temperature dependent
electrical conductivity of bilayer graphene in the presence of
long-range charged impurity scattering as well as an additional short-range
disorder of independent origin, establishing that both scattering
mechanisms contribute significantly to determining bilayer transport
properties. We find that although strong screening properties of
bilayer graphene lead to qualitative differences with the
corresponding single layer situation, both systems exhibit the
linearly density dependent conductivity at high density and the
minimum graphene conductivity behavior around the charge neutrality
point due to the formation of inhomogeneous electron-hole puddles
induced by the random charged impurity centers.
\end{abstract}
\maketitle


Ever since the discovery of graphene \cite{Geim1} its transport
properties as functions of carrier density and temperature have been
of key fundamental and technological interest. The fundamental
interest arises from the unique linear massless chiral Dirac
dispersion of electrons (holes) in the graphene conduction (valence)
band with the system being a zero-gap semiconductor which on doping
(or external field induced gating) changes its character continuously
from being an `electron-metal' to a `hole-metal' as it goes through
the charge neutral Dirac point. Unique transport properties of
massless, gapless, and chiral Dirac particles in 2D single layer
graphene (SLG) as functions of their density and temperature have
attracted a great deal of experimental and theoretical attention over
the last few years. 

More recently, however, carrier transport in 2D
bilayer graphene (BLG) has attracted considerable attention
\cite{morozov2008,xiao2009,bilayer}.  In BLG,
the carriers tunnel quantum mechanically between the two layers
leading to a modified band dispersion which is approximately parabolic
with an effective mass of about 0.033$m_e$ \cite{bilayer}. BLG transport thus
involves dynamics of chiral, parabolic dispersion carriers in the zero
band gap situation in contrast to the chiral, linear dispersion Dirac
carrier system for SLG. In addition to the considerable fundamental
interest in BLG transport, there is a great deal of technological
interest in understanding BLG transport because of the possibility of
opening a band gap in BLG via an external electric field \cite{bilayer}. The
existence of such a band gap coupled with the ability to tune the
carrier density using an external gate voltage leads to the possibility of
graphene-based transistor devices with obvious implications for
microelectronic and nanoelectronic applications.

In this Letter we develop the first comprehensive and quantitative
theory for BLG carrier transport in the presence of random charged
impurity (i.e., long-range Coulomb \cite{Adam2008}) and short-range
scattering \cite{Katsnelson}. Coulomb scattering has 
emerged as the leading, if not the dominant, scattering mechanism
controlling SLG carrier transport. In particular, the observed linear
dependence of the SLG conductivity on carrier density away from the
charge neutral Dirac point as well as the observed weak temperature
dependence of conductivity are both naturally explained as arising
from the screened Coulomb scattering of chiral Dirac carriers by
random charged impurities invariably present in the graphene
environment \cite{hwang2007}. 
Although the precise quantitative importance of charged
impurity scattering in determining all aspects of SLG transport
properties is still occasionally debated \cite{debate} in the
literature, there is 
wide consensus that SLG transport both around and away from the Dirac
point is determined mostly, if not entirely, by screened Coulomb
scattering from random charged impurity centers. It
is, therefore, 
reasonable to assume that charged impurity scattering also dominates
BLG transport, and to investigate theoretically the effect of charged
impurity scattering on BLG conductivity. 

We start by providing a qualitative conceptual discussion of BLG
transport properties vis a vis SLG transport. Because of the linear
(SLG) versus quadratic (BLG) energy dispersion in the two cases, the
density of states is linear (constant) in SLG (BLG), leading
to very distinct screening behavior in two cases
\cite{Hwang_SLG,Hwang_BLG}: 
$q_{TF}^{SLG} 
\propto k_F$; $q_{TF}^{BLG} \propto k_F^0 =$constant, where $q_{TF}$ is
the long wavelength Thomas-Fermi screening wave vector and
$k_F\propto \sqrt{n}$ is the Fermi wave vector (with $n$ as the
tunable 2D carrier density). Therefore, the screened Coulomb impurity
potential, $u_i(q)$, behaves very differently in the two systems:
$u_i^{SLG}(q) \sim k_F^{-1}$; $u_i^{BLG} \sim (q_{TF}+k_F)^{-1}$, where
$q_{TF} \equiv q_{TF}^{BLG} = (4me^2)/(\kappa \hbar^2) \approx
10^7$cm$^{-1}$, where $\kappa$ is the background environmental
dielectric constant. Thus, the nature of screened Coulomb scattering
is qualitatively different in the two systems --- in fact, the
screened Coulomb disorder in the SLG behaves as unscreened Coulomb
interaction (i.e., $v(q) \sim 1/q$) \cite{Hwang_SLG} whereas the screened 
Coulomb disorder in the BLG behaves similar to the 2D screened Coulomb
interaction although there are some differences \cite{Hwang_BLG}. It
is, therefore, quite puzzling that 
experimentally the 
two systems have very similar carrier transport properties
\cite{morozov2008,xiao2009}.  Away from
the charge neutral point, CNP, ($n=0$), both manifest a conductivity,
$\sigma$, approximately proportional to $n$,
whereas for $n\sim0$ 
the conductivity is approximately constant ($\sigma \sim 1-10$ 
$e^2/h$), forming the much-discussed graphene minimum conductivity
plateau in both cases. The striking similarity of the experimentally
observed density dependent conductivity $\sigma(n)$ in the two systems
presents a serious challenge to transport theories since the nature of
screened Coulomb disorder in the two systems is fundamentally
different.

In this Letter we propose that BLG carrier transport is controlled
almost equally by two distinct and independent physical scattering
mechanisms: Screened Coulomb disorder due to random
charged impurities in the environment and a short-range disorder (of
unknown origin) which is important throughout (i.e. all the way from
the CNP manifesting the plateau-like minimum
conductivity behavior to the high-density conductivity increasing
linearly with carrier density). We find that neither pure Coulomb
disorder nor pure short-range disorder by itself can
explain {\it both} the density and the $T$ dependence of the
experimentally observed BLG conductivity behavior. 
Earlier theoretical work on BLG transport \cite{Adam2008},
considering only screened Coulomb disorder, included drastic
approximations (e.g. complete screening) which are
both unreliable and uncontrolled. We emphasize that the actual amount
of Coulomb and short-range disorder we need to qualitatively and
semi-quantitatively explain the existing BLG transport data is
comparable to that already used in understanding the SLG transport
properties, and as such, there is no arbitrary data fitting procedure
in our theory. The new element of physics, not appreciated 
in the earlier literature, is that BLG screening is
quantitatively much stronger than SLG screening, rendering the effect
of Coulomb scattering relatively less important in BLG (compared
with SLG). This then leads to any other scattering mechanism present
in the system to become much more important for BLG than for SLG, and
hence short-range scattering is much more significant for BLG
than for SLG. This is easily seen by noting that the ratio `$r$' of
screened Coulomb disorder in SLG to BLG goes as $r\equiv
u_i^{SLG}/u_i^{BLG} \equiv (1+q_{TF}/k_F) \approx
(1+11.5/\sqrt{\tilde{n}})$ where $\tilde{n} = (n/10^{12})$. Thus,
$r \gg 1$ in the typical graphene experimental carrier density regime.

The density and temperature dependent  BLG conductivity is given
within the Boltzmann transport theory by
$\sigma = ({e^2}/{m}) \int d\epsilon D(\epsilon) \epsilon
\tau(\epsilon) \left (    
  -{\partial f}/{\partial \epsilon} \right ),$
where $D(\epsilon)$, $\tau(\epsilon)$, $f$ are respectively the energy
dependent density of states, the transport relaxation time (which
depends explicitly on the scattering mechanism), and the
(finite-T) Fermi distribution function. $\tau$ is given by
\begin{eqnarray}
\frac{1}{\tau(\epsilon_{s\vk})} = \frac{2\pi n_0}{\hbar} \int \frac{d^2
  k'}{(2\pi)^2} | \langle V_{s\vk,s\vk'}\rangle |^2 g(\theta_{\vk
  \vk'}) \nonumber \\ 
\times [1-\cos\theta_{\vk\vk'}]
\delta\left (\epsilon_{s\vk} - \epsilon_{s\vk'} \right ),
\label{tau}
\end{eqnarray}
where $\epsilon_{s\vk}=s\hbar^2k^2/2m$ is the carrier energy for the
spin/pseudospin state `$s$' and 2D wave vector $\vk$,
$\langle V_{s\vk,s\vk'} \rangle$ is the matrix element of the
appropriate disorder 
potential, $g(\theta)=[1+\cos(2\theta)]/2$ is a wave function form
factor associated with the chiral matrix of graphene (and is
determined by the BLG band structure). In Eq.~(\ref{tau}), $n_0$ is the
appropriate 2D areal concentration of the impurity centers giving rise
to the random disorder potential $V$. We consider two different kinds
of disorder scattering: ($i$) screened Coulomb disorder $u_i$, and
($ii$) short-range disorder $V_0$, writing
$n_0| \langle V_{s\vk,s\vk'} \rangle |^2 =  n_i |u_i|^2= n_i|
{v_i(q)}/{\epsilon(q)} |^2$ and 
$n_0| \langle V_{s\vk,s\vk'} \rangle |^2
= n_d V_0^2$, respectively.
Here $q\equiv |\vk-\vk'|$, $v_i(q) = (2\pi e^2/\kappa q)e^{-qd}$ is
the 2D Coulomb potential,
and $V_0$  a constant short-range (i.e. a
$\delta$-function in real space) potential. (Note that we use $n_i$ to
denote the charged impurity density.) 
In the expression for $v_i$, $d$ is the average distance of the
charged impurities from BLG. We find that in contrast to SLG, in BLG
the conductivity 
is essentially independent of $d$ for $d<10\;\AA$ due to the short
BLG screening length.
The dielectric screening
function $\epsilon(q)$ entering the effective screened Coulomb
disorder, which depends on both $q$ and
$T$, was calculated in Ref.~\cite{Hwang_BLG} at
$T=0$, and we 
have now generalized it to $T\neq 0$ as shown in Fig.~1. Note that BLG
screening peaks at $2k_F$ which is also the important BLG scattering
wave vector due to the form factor $g$ in Eq.~(1).

\begin{figure}
\epsfysize=2.3in
\epsffile{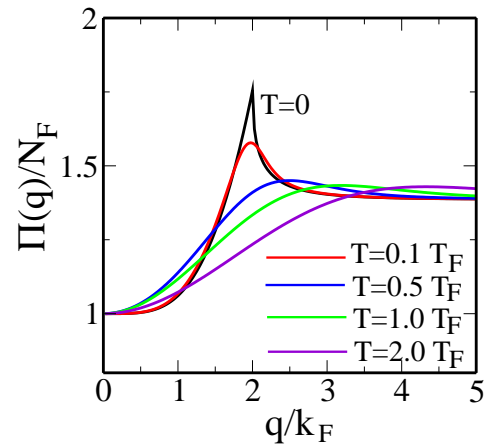}
\caption{ (Color online)
The wave vector dependent 2D BLG polarizability $\Pi$ (in units of the
density of states at the Fermi level $N_F$) for different
values of $T/T_F=$0, 0.1, 0.5, 1.0, 2.0, where $T_F=E_F/k_B$ is the
Fermi temperature ($T_F \approx 831$K for $n=10^{12}$ cm$^{-2}$) and
the dielectric function is given by $\epsilon = 1+ v_c \Pi$ where
$v_c$ is the 2D Coulomb interaction. The ordinate can be taken as a
measure of the strength of BLG screening in units of Thomas-Fermi
screening constant $q_{TF} \approx 10^7$cm$^{-1}$.
\label{fig:polariz}
}
\end{figure}

It is straightforward to calculate the analytical density dependence 
(for $T/T_F\ll 1$) of BLG conductivity from the above formulae:
$ \sigma(n) \sim n^2$ for unscreened Coulomb disorder;
$ \sigma(n) \sim n$ for  overscreened Coulomb disorder;
$ \sigma(n) \sim n^\alpha,\; 1<\alpha<2$ for screened Coulomb disorder;
$ \sigma(n) \sim n$ for short range disorder.
We emphasize that $\alpha(n)$ is a density dependent exponent which
varies slowly changing from 1 at low density to 2  
at high density --  in the BLG experimental density range $\alpha\approx 1.2$.
Increasing temperature, in general, suppresses screening, leading to 
a slight enhancement of the exponent $\alpha$. 

It is obvious from the above discussion that Coulomb disorder by itself cannot
explain the experimentally observed linear density dependence of
$\sigma(n)$ in BLG. 
By contrast, the SLG conductivity for Coulomb disorder, either screened or 
unscreened, follows the linear $\sigma(n)\sim n$ behavior whereas the short
range disorder leads to a density independent $\sigma(n)\sim n ^0$
behavior. 
Thus, the carrier transport physics is fundamentally and
qualitatively different in SLG and BLG: SLG is
always dominated by Coulomb disorder 
except at extreme high densities, whereas in BLG the short-range disorder is 
always equally, if not more, important for all densities.

%
\begin{figure}
\includegraphics[width=6.5cm]{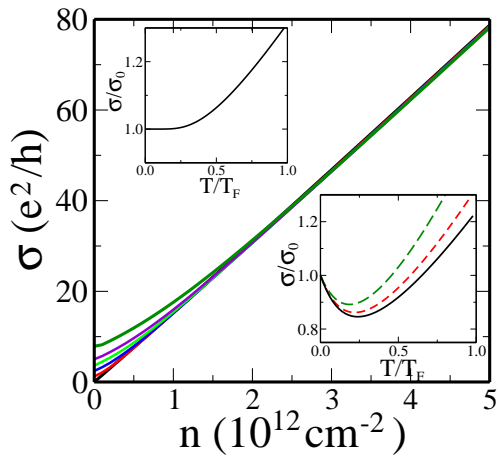}
\caption{(Color online) (a) Density dependence of BLG
  conductivity for different 
  temperatures, T=0, 50, 100, 150, 200, 300 K (from bottom to top),
   $n_i = 10^{11}cm^{-2}$ and
   $n_dV_0^2 = 2.0 (eV \AA)^2$. 
Top inset shows $\sigma$ as a function of $T$ in presence of short
range disorder. The scaled conductivity applies for all densities.
Bottom inset shows $\sigma$ as a function of $T$ in presence of
screened Coulomb disorder for different densities
$n=5$, 10, 30$\times 10^{11}cm^{-2}$ (from bottom to top). 
%
\label{fig:sigma_boltz}
}
\end{figure}
%

In  Fig.~\ref{fig:sigma_boltz} we present our full numerically calculated
finite $T$ BLG conductivity as a function of $n$ for
reasonable representative values of disorder strength.
The results shown in Fig.~\ref{fig:sigma_boltz} are in good agreement
with recent experimental results \cite{morozov2008,xiao2009}, but we
do not make any attempt to obtain 
exact quantitative agreement since the experimental results show substantial
sample-to-sample variations \cite{geim_pc}. Instead we discuss  the 
salient qualitative features of our results:
({\it i}) The calculated density dependence is approximately linear over 
a wide density range as seen experimentally;
({\it ii}) the temperature dependence is very weak at higher densities as
observed in recent experiments \cite{morozov2008};
({\it iii}) at low densities, where $T/T_F$ is not too small, there is a strong
insulating-type $T$ dependence arising from the thermal excitation
of carriers (which is exponentially suppressed at higher densities) and 
energy averaging, as observed experimentally \cite{morozov2008};
({\it iv}) when the dimensionless temperature is very small $(T/T_F\ll 1)$
our theory necessarily predicts (see lower inset of Fig.~\ref{fig:sigma_boltz})
a weak linear-in $T$ metallic $T$ dependence arising from the 
temperature dependence of the screened charges impurity scattering, i.e. 
the thermal suppression of the $2k_F$-peak associated with back-scattering
in Fig.~\ref{fig:polariz} -- this effect is enhanced in BLG due to
the importance of backscattering whereas it is suppressed in SLG.
By contrast, for the short-range
disorder $\sigma$ always increases with $T$, as shown in the upper
inset of Fig.~\ref{fig:sigma_boltz}. 

While the calculations for Fig.~\ref{fig:sigma_boltz} are all carried out
for graphene on ${\rm SiO_2}$ substrate (corresponding to $\kappa\approx 2.5$),
we have carried out detailed calculations of $\sigma(n,T,\kappa)$ as a function
of the effective background dielectric constant (not shown) since 
screened Coulomb disorder should
manifest a strong 
dependence on $\kappa$. Our calculations show the $\kappa$-dependence of the 
net conductivity to be almost unobservable -- for example, 
covering BLG on SiO$_2$ with ice,
thus changing $\kappa$ from 2.5 to 3.5, would
only increase the conductivity by 1\%, for $n_i=10^{11}{\rm cm}^{-2}$,
and 10\%, for $n_i=10^{12}{\rm cm}^{-2}$. 

\begin{figure}[tb]
 \begin{center}
  \includegraphics[width=8.5cm]{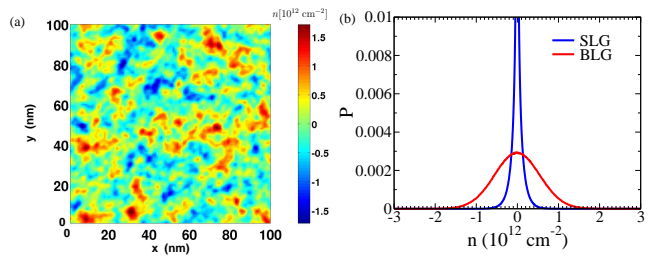} 
  \caption{ 
           (Color online). 
           (a) $n(\rr)$ of BLG at the CNP
           for a single 
           disorder realization with $n_i=10^{11}{\rm cm}^{-2}$ and $d=1$~nm.
           (b) Disorder averaged $P(n)$,
           at the CNP for BLG (SLG) red (blue) for
           $n_i=10^{11}{\rm cm}^{-2}$ and $d=1$~nm. 
           For SLG $P(n=0)\approx 0.1$, out of scale. 
           The corresponding $n_{\rm rms}$ is $5.5\;10^{11}{\rm
           cm}^{-2}$ for BLG 
           and $1.2\;10^{11}{\rm cm}^{-2}$ for SLG. 
          } 
  \label{fig:n_land}
 \end{center}
\end{figure} 

The Boltzmann theory presented so far assumes a completely homogeneous carrier
density landscape over the BLG sample, and leads to $\sigma(T=0)=0$
at the CNP $(n=0)$. It is, however, well known that
SLG breaks up into an inhomogeneous landscape of electron-hole puddles
around the CNP \cite{martin2008, rossi2008, zhang2009}.
We expect the same physics of electron-hole puddles to dominate the BLG 
properties around the CNP, and this will give rise to
a finite ``minimum conductivity'', $\sigma_{\rm min}\equiv \sigma(n=0)$
even in the limit $T\to 0$.
We have investigated the properties of BLG puddle formation by calculating
the ground state of the BLG in the presence of random charged impurities
solving the density functional equations within a local density
approximation. \cite{rossi2008}
Using the resultant ground state density landscape
$n({\bf r})$ 
with $n\equiv \langle n(\rr)\rangle_\rr$ and averaging over
disorder realizations we have then calculated the BLG
transport properties using the effective medium theory, EMT~\cite{rossi2009}.
The details of the calculations will be presented elsewhere, here we present
the realistic conductivity results in the puddle dominated regime 
obtained using our EMT.

In Fig.~\ref{fig:n_land}~(a) we show our calculated zero-density
puddle structure for the BLG for a single disorder realization with
$n_i=10^{11}{\rm cm}^{-2}$. 
We note that by neglecting the disorder induced inhomogeneities 
BLG (SLG) would have perfect (vanishing) linear screening properties
as $n\to 0$,
whereas the strong
carrier density fluctuations and associated electron-hole puddle structure
induced by the charged impurities close to the CNP 
are qualitatively very similar in BLG and SLG.
The linear (SLG) versus parabolic (BLG) carrier dispersion has 
the quantitative effect of modifying the form of the probability distribution $P(n)$
with the result that $P(n)$ is much wider in BLG than in SLG, as shown in Fig.~\ref{fig:n_land}~(b).

%
\begin{figure}[tb]
 \begin{center}
  \includegraphics[width=6.5cm]{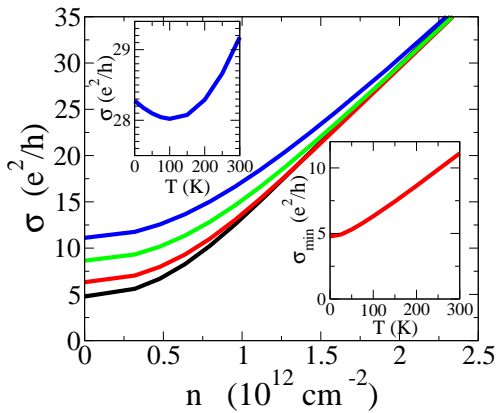}
  \caption{
           (Color online). 
           Conductivity as a function of $n$ obtained using the EMT
           for $n_i=10^{11}{\rm cm}^{-2}$, $n_dV_0^2=2\;({\rm
           eV\AA})^2$ and several values of 
           $T$, from top to bottom: $T=300,\;200,\;100,\;0$.
           From the slope for $n>10^{12}{\rm cm}^{-2}$ for $T=0$ we extract
           a mobility of $4000$~cm$^2$/Vs.
           In the lower (upper) inset $\sigma_{\rm min}$
           ($\sigma(n=1.9\;10^{12}{\rm cm}^{-2})$) as a function of $T$ for the
           same values of disorder strengths. 
          } 
  \label{fig:sigma_emt}
 \end{center}
\end{figure} 

Finally in Fig.~\ref{fig:sigma_emt} we show $\sigma(n)$ for BLG 
for several values of $T$, taking into account  
the inhomogeneity of the puddles.
These results are calculated numerically through the EMT by combining the
density functional electronic structure calculation with the full
Boltzmann transport 
theory. At high density $(\gg n_i)$, the theory gives the same results
as that obtained from the Boltzmann theory in the homogeneous case
(i.e.~Fig.~\ref{fig:sigma_boltz}), 
but at low densities there are significant deviations. For example, we
now get a well-defined 
finite $\sigma_{min}$ even at $T=0$.
We show the $T$ dependence of the conductivity
for $n=0$ and $n\gg n_i$  in the insets obtaining good agreement with
the experimental finding of a strong insulating $T$
dependence of $\sigma_{\rm min}$. Our calculated $\sigma_{\rm min}$
depends weakly on $n_i$ with no universally
discernible functional dependence on $n_i$.
For the range of values of $n$ and $n_i$ considered
using a more realistic hyperbolic BLG band dispersion
in the theory does not change our results and conclusions in any qualitative
manner. A small BLG band gap ($\ll E_F$) at $k=0$ also has no
significant effect.
%


We conclude by emphasizing the similarity and the difference
between  BLG and SLG transport from the perspective of our
transport theory considerations. We find that both manifest
a non-universal sample dependent minimum conductivity
at the CNP arising from the electron-hole
puddle formation due to the 
presence of long-range Coulomb disorder. 
We find that BLG manifests a fairly strong insulating
temperature dependence of $\sigma_{\rm min}$, which is consistent
with experimental observations. 
We find that at high
density both BLG and SLG manifest a linearly increasing conductivity
with increasing carrier density, as observed experimentally.
However the physical origin for this linear dependence 
is quite different in the two systems: while in the SLG
this linearity arises entirely from Coulomb disorder, 
in the BLG we must invoke the existence of a short-range
disorder scattering to explain the linearity.
The importance of short-range scattering in BLG
compared with SLG is understandable based on BLG 
screening being much stronger than SLG screening.
The existence of the short-range disorder 
naturally explains the strong insulating temperature
dependence of BLG conductivity since short-range
disorder leads to a $\sigma(T)$ increasing with $T$
for all $T$, whereas Coulomb scattering always leads to
a $\sigma(T)$ decreasing (linearly at first) with increasing
$T$ for $T\ll T_F$.
A possible direct way of experimentally investigating the competing
effects of short-range versus Coulomb scattering in BLG will be to
plot the measured low-temperature conductivity against the carrier
density on a log-log plot to see whether the power law exponent at
higher densities is  larger 
than unity, thus
implying the dominance of Coulomb 
scattering. 
The origin of the strong short-range disorder in BLG
has to be considered unknown at this stage. It could be neutral
defects such as lattice vacancies (only a minuscule concentration
of $10^{-7}$ or less would be needed), or ripples, or resonant
short-range scattering, or some other mechanism.

Work supported by ONR-MURI and NRI-NSF-SWAN.



\end{document}